# Birefringent Majorana fermions

Hu Zhang

Key Laboratory of Optic-Electronic Information and Materials of Hebei Province, Institute of Life Science and Green Development, College of Physics Science and Technology, Hebei University, Baoding 071002, P. R. China

E-mails: zhanghu@hbu.edu.cn

We propose the concept of birefringent Majorana fermions as collective excitations of quantum interacting many-body states motivated by the already known quasiparticles including Dirac fermions, birefringent Dirac fermions and Majorana fermions. Low energy effective models are constructed to understand the basic mathematical structures of equations of motion and physical properties. We find that the equation of motion, with two different "speed of light", describing birefringent Majorana fermions in condensed matter physics is the birefringence version of the original Majorana equation of real relativistic fermions.

The concept of quasiparticle in condensed matter physics is of great value to understand the physical properties of various quantum systems. One of typical examples is the massless Dirac fermions which are the low-energy excitations of some solids (known as Dirac semimetals) such as graphene which is a two-dimensional (2D) material with the honeycomb structure [1]. The linear electronic dispersion around the Dirac point is plotted in Fig. 1(a). Due to such unusual Dirac-like electronic excitations, graphene has great importance both in technological applications and fundamental scientific interest. As the three-dimensional (3D) analog of graphene, Dirac semimetal $Na_3Bi$ was predicted by the first-principles calculations and then confirmed experimentally [2,3]. These bulk 3D Dirac points (fourfold nodal points) are protected by the crystal symmetry. There exist nontrivial Fermi arcs on the surfaces of Dirac semimetals. In 2011, a lattice model for fermions was found to have low-energy excitations describing by the gapless birefringent Dirac fermions in which the doubly degenerate Dirac cones split into birefringent cones with different "speeds of light" [4]. The linear electronic dispersion of the gapless birefringent Dirac fermions is plotted in Fig. 1(b). The broken chiral symmetry is important in this model. Soon after this, gapless birefringent Dirac fermions were predicted in the 3D material CaAgAs (nonmagnetic) [5,6]. Different from the centrosymmetric Dirac semimetals such as $Na_3Bi$, CaAgAs has the $P6_3mc$ polar space group in which inversion symmetry is broken. To exhibit the polar symmetry (usually appears in ferroelectrics), this type of Dirac semimetals was called polar Dirac semimetals. Dirac points in this polar Dirac semimetal are protected by the sixfold rotational symmetry. Due to the broken of inversion symmetry, electronic bands that do not along the rotational axis are split by spin-orbit coupling (SOC), which results in birefringent Dirac cones.

In particle physics, a Dirac fermion is a spin-1/2 particle which is distinct from its antiparticle described by the Dirac equation. The solution of the Majorana equation describes a Majorana fermion that is a spin-1/2 neutral particle, but with no distinction between antiparticle and particle [7]. In condensed matter physics, Majorana fermions can occur in some many-body systems as emergent quasiparticles. One example is the gapless Majorana fermion surface state of topological superfluid $^3$He-B (time reversal invariant) [8]. In addition, topological superconductors were considered to have gapless surface states consisting of Majorana fermions [9]. In Fig. 1(c), the quasiparticle spectra of Majorana fermions are shown, in which dashed lines indicate that the negative energy part ($E < 0$) are redundant. The experimental discovery and potential applications of the Majorana fermions in condensed matter physics remain an inspiration.

From the development of the concept of Dirac fermions to birefringent Dirac fermions in condensed matter physics, it is natural to propose the concept of massless birefringent Majorana fermions by analogy. In Fig. 1(d), the quasiparticle spectra of birefringent Majorana fermions are given. From Fig. 1, we may consider that the birefringent Majorana fermions can be viewed as the Majorana version of birefringent Dirac fermions or the birefringent version of Majorana fermions. In this work, we will establish the fundamental physics of massless birefringent Majorana fermions with the help of low energy effective models.

In quantum mechanics, we have the following full space-time equation ($\hbar = 1$)

$$i\partial_t \Phi = \mathcal{H}\Phi. \tag{1}$$

We can separate time with the usual phase *Ansatz* and thus find out the energy eigenvalue spectrum [10]

$$\Phi = e^{-iEt}\Phi_E, \mathcal{H}\Phi_E = E\Phi_E. \tag{2}$$

With the help of the Pauli matrices

$$\sigma^1 = \begin{pmatrix} 0 & 1 \\ 1 & 0 \end{pmatrix}, \sigma^2 = \begin{pmatrix} 0 & -i \\ i & 0 \end{pmatrix}, \sigma^3 = \begin{pmatrix} 1 & 0 \\ 0 & -1 \end{pmatrix}, \tag{3}$$

we can write down a Hamiltonian describing the low energy physics of the massless Dirac fermions in condensed matter physics [2]

$$\mathcal{H}_1 = v_F \begin{pmatrix} \sigma^1 p_x + \sigma^2 p_y & 0 \\ 0 & -(\sigma^1 p_x + \sigma^2 p_y) \end{pmatrix}, \tag{4}$$

where $v_F$ is the Fermi velocity and $p_{x,y} = -i\partial_{x,y}$. This Hamiltonian has the same form as the Hamiltonian in Dirac equation in the Weyl representation. In the momentum space with an $e^{i\mathbf{k}\cdot\mathbf{r}}$ *Ansatz*, the energy eigenvalue is

$$E = \pm v_F \sqrt{k_x^2 + k_y^2}, \tag{5}$$

where $k$ is measured relatively to Dirac points. Each energy band is doubly degenerate because of the coexistence of inversion symmetry and time reversal symmetry. This gives a relativistic dispersion in which $v_F$ plays the role of the "speed of light", as shown in Fig. 1(a). There are doubly degenerate Dirac cones in the $k_x$-$k_y$ plane.

One the other hand, according to previous works [5,6], the gapless birefringent Dirac fermions in CaAgAs (the rotational axis is set along the $z$ direction) can be described by the Hamiltonian

$$\mathcal{H}_2 = \begin{pmatrix} v_1(\sigma^2 p_x - \sigma^1 p_y) & 0 \\ 0 & -v_2(\sigma^2 p_x - \sigma^1 p_y) \end{pmatrix}. \tag{6}$$

where $v_1$ and $v_2$ are materials-dependent real model parameters with $v_1 \neq v_2$. Remarkably, this Hamiltonian is the double copy of the Rashba term. The energy eigenvalue is

$$E = \pm v_{1,2}\sqrt{k_x^2 + k_y^2}. \tag{7}$$

These are relativistic dispersions with $v_{1,2}$ playing the role of different "speed of light" as shown in Fig. 1(b). In the $k_x$-$k_y$ plane, there exist birefringent Dirac cones with distinct slopes because of broken inversion symmetry, which is analogous to birefringence of light in crystals (such as calcite). In a word, the intertwined broken inversion symmetry, time reversal invariant, SOC and rotational symmetry lead to the emergence of birefringent Dirac fermions in CaAgAs.

Different from the Dirac equation, the Majorana equation (Eq. (1) in the Majorana representation) is real and thus has a real solution [7,10]. Then the time-dependent wave function $\Phi$ satisfies a reality condition

$$\Phi^* = \Phi. \tag{8}$$

In this case, $\mathcal{H}$ in Eq. (1) is imaginary and thus possesses the conjugation symmetry

$$\mathcal{H} = -\mathcal{H}^*. \tag{9}$$

As a result, to each positive energy eigenmode there corresponds a negative energy mode

$$\Phi^*_{+E} = \Phi_{-E}. \tag{10}$$

A quantum field can be constructed by superposing the energy eigenmodes $\Phi_E$ with annihilation and creation quantum operators. Under the reality condition, the Majorana field operator [10] (usually denoted as $\widehat{\Psi}$) is

$$\widehat{\Psi} = \sum_{E>0} (a_E e^{-iEt}\Phi_E + a_E^\dagger e^{iEt}\Phi_E^*), \tag{11}$$

where the $a_E^\dagger$ operator creates negative energy excitations and the $a_E$ operator annihilates positive energy excitations. For the conventional anticommutation algebra of $a_E$ and $a_E^\dagger$ we obtain

$$\widehat{\Psi}_a^\dagger = \widehat{\Psi}_a. \tag{12}$$

This reality condition demonstrates that a Majorana particle state (quasiparticle) is identified with its antiparticle state (quasihole) at the level of Majorana field operators.

Now we construct the low energy effective Hamiltonian describing the massless birefringent Majorana fermions as collective excitations of some yet unknown quantum interacting many-body states. These models can help us to understand the basic mathematical structures of equations and physical properties for the massless birefringent Majorana fermions, which is the main aim of this work. One may observe birefringent Majorana fermions in a superconductor, or an interacting many-body system whose low energy effective description is that of a superconductor. We start from the five $4 \times 4$ Dirac $\Gamma$ matrices defined in Ref. [11]

$$\Gamma^1 = \sigma^1 \otimes \tau^1, \Gamma^2 = \sigma^2 \otimes \tau^1, \Gamma^3 = \sigma^3 \otimes \tau^1,$$
$$\Gamma^4 = I \otimes \tau^2, \Gamma^5 = I \otimes \tau^3. \tag{13}$$

Here $\tau^{1,2,3}$ are another set of Pauli matrices and I is the $2 \times 2$ identity matrix. These Dirac matrices satisfy Clifford algebra $\{\Gamma^a, \Gamma^b\} = 2\delta_{ab}$. The other ten Dirac $\Gamma$ matrices are given by $\Gamma^{ab} = [\Gamma^a, \Gamma^b]/2i$. We note that $\Gamma^3$ and $\Gamma^5$ are block diagonal real matrices

$$\Gamma^3 = \begin{pmatrix} 0 & 1 & 0 & 0 \\ 1 & 0 & 0 & 0 \\ 0 & 0 & 0 & -1 \\ 0 & 0 & -1 & 0 \end{pmatrix}, \Gamma^5 = \begin{pmatrix} 1 & 0 & 0 & 0 \\ 0 & -1 & 0 & 0 \\ 0 & 0 & 1 & 0 \\ 0 & 0 & 0 & -1 \end{pmatrix}. \tag{14}$$

To our purpose, we can modify them as

$$\tilde{\Gamma}^3 = \left(\frac{\alpha_1+\alpha_2}{2}\Gamma^3 + \frac{\alpha_1-\alpha_2}{2}\Gamma^{45}\right) = \begin{pmatrix} 0 & \alpha_1 & 0 & 0 \\ \alpha_1 & 0 & 0 & 0 \\ 0 & 0 & 0 & -\alpha_2 \\ 0 & 0 & -\alpha_2 & 0 \end{pmatrix},$$

$$\tilde{\Gamma}^5 = \left(\frac{\alpha_1+\alpha_2}{2}\Gamma^5 + \frac{\alpha_1-\alpha_2}{2}\Gamma^{34}\right) = \begin{pmatrix} \alpha_1 & 0 & 0 & 0 \\ 0 & -\alpha_1 & 0 & 0 \\ 0 & 0 & \alpha_2 & 0 \\ 0 & 0 & 0 & -\alpha_2 \end{pmatrix}, \tag{15}$$

which also satisfies Clifford algebra $\{\tilde{\Gamma}^3, \tilde{\Gamma}^5\} = 0$. Here $\alpha_{1,2}$ are real parameters. Based on these two matrices, we can write down the Hamiltonian

$$\mathcal{H}_3 = \tilde{\Gamma}^3 p_x + \tilde{\Gamma}^5 p_y = \begin{pmatrix} \alpha_1 p_y & \alpha_1 p_x & 0 & 0 \\ \alpha_1 p_x & -\alpha_1 p_y & 0 & 0 \\ 0 & 0 & \alpha_2 p_y & -\alpha_2 p_x \\ 0 & 0 & -\alpha_2 p_x & -\alpha_2 p_y \end{pmatrix}. \tag{16}$$

Clearly, $\mathcal{H}_3$ is imaginary and thus the full space-time equation (1) is real, as required to describe Majorana fermions. The corresponding energy eigenvalue is

$$E = \pm \alpha_{1,2}\sqrt{k_x^2 + k_y^2}. \tag{17}$$

There are two negative energy solutions and two positive energy solutions. These are relativistic dispersions, as shown in Fig. 1(d), with $\alpha_{1,2}$ playing the role of different "speed of light". Therefore, $\mathcal{H}_3$ is a low energy effective Hamiltonian describing the massless birefringent Majorana fermions. The quantum operator $\widehat{\Psi}$ is

$$\widehat{\Psi}(t,\boldsymbol{r}) = \sum_n \int \frac{d^2k}{(2\pi)^2} \{a_n(\boldsymbol{k})e^{-i(E_n t - \boldsymbol{k}\cdot\boldsymbol{r})}\Phi_n(\boldsymbol{k}) + a_n^\dagger(\boldsymbol{k})e^{i(E_n t - \boldsymbol{k}\cdot\boldsymbol{r})}\Phi_n^*(\boldsymbol{k})\}, \tag{18}$$

where $n = 1, 2$ corresponding to the eigenvalue $E$ with different "speed of light" $\alpha_1$ and $\alpha_2$.

Similarly, we can modify the other two real matrices $\Gamma^{14}$ and $\Gamma^{45}$ as

$$\widetilde{\Gamma}^{14} = \left(\frac{\beta_1+\beta_2}{2}\Gamma^{14} + \frac{\beta_1-\beta_2}{2}\Gamma^{23}\right) = \begin{pmatrix} 0 & 0 & \beta_1 & 0 \\ 0 & 0 & 0 & -\beta_2 \\ \beta_1 & 0 & 0 & 0 \\ 0 & -\beta_2 & 0 & 0 \end{pmatrix},$$

$$\widetilde{\Gamma}^{45} = \left(\frac{\beta_1+\beta_2}{2}\Gamma^{45} + \frac{\beta_1-\beta_2}{2}\Gamma^{3}\right) = \begin{pmatrix} 0 & \beta_1 & 0 & 0 \\ \beta_1 & 0 & 0 & 0 \\ 0 & 0 & 0 & \beta_2 \\ 0 & 0 & \beta_2 & 0 \end{pmatrix}. \tag{19}$$

We should note that $\{\widetilde{\Gamma}^{14}, \widetilde{\Gamma}^{45}\} \neq 0$ different from the above case. Again, $\beta_{1,2}$ are real parameters. Based on these two matrices, we can write down the Hamiltonian

$$\mathcal{H}_4 = \widetilde{\Gamma}^{14}p_x + \widetilde{\Gamma}^{45}p_y = \begin{pmatrix} 0 & \beta_1 p_y & \beta_1 p_x & 0 \\ \beta_1 p_y & 0 & 0 & -\beta_2 p_x \\ \beta_1 p_x & 0 & 0 & \beta_2 p_y \\ 0 & -\beta_2 p_x & \beta_2 p_y & 0 \end{pmatrix}. \tag{20}$$

Clearly, $\mathcal{H}_4$ is also imaginary as required. This Hamiltonian also appears in Ref. but with different physical meaning. The corresponding energy eigenvalue is

$$E = \pm \beta_{1,2}\sqrt{k_x^2 + k_y^2}. \tag{21}$$

These relativistic dispersions are similar with those of $\mathcal{H}_3$. However, $\mathcal{H}_3$ and $\mathcal{H}_4$ have very different eigenvectors and thus they describe birefringent Majorana fermions with different physical properties.

Finally, it is helpful to compare the full space-time equation of motion for the massless birefringent Majorana fermions in condensed matter physics and the equation in Majorana's original formulation of real relativistic fermions. Obviously, $i\mathcal{H}$ is real if $\mathcal{H}$ is imaginary. From Eq. (1) we have

$$(\partial_t + i\mathcal{H})\Phi=0. \tag{22}$$

For $\mathcal{H}_3$ we have

$$[\partial_t + i(\tilde{\Gamma}^3 p_x + \tilde{\Gamma}^5 p_y)]\Phi=0, \tag{23}$$

or written in a clear form

$$[\partial_t + (\tilde{\Gamma}^3 \partial_x + \tilde{\Gamma}^5 \partial_y)]\Phi=0. \tag{24}$$

This equation of motion is purely real as required. On the other hand, the equation in Majorana's original formulation of real relativistic fermions is (in the case of massless)

$$[\partial_t - c(\boldsymbol{\alpha}, \text{grad})]U=0, \tag{25}$$

where $\alpha_{x,y,z}$ are Dirac matrices in Majorana representation (real), $\text{grad} = (\partial_x, \partial_y, \partial_z)$, $c$ is the speed of light and U is used by Majorana [12]. The equations of motion (23) and (24) have exactly the same form except that there are two different "speed of light" $\alpha_1$ and $\alpha_2$ in the equation (23) contained in $\tilde{\Gamma}^3$ and $\tilde{\Gamma}^5$. We can say that the equation of motion (23) describing massless birefringent Majorana fermions in condensed matter physics is the 2D birefringence version of the Majorana equation of real relativistic fermions. The corresponding energy eigenvalue (17) can be compared with Einstein's energy-momentum relation (massless)

$$E^2 = p^2 c^2. \tag{26}$$

In summary, we have proposed the concept of a spin-1/2 quasiparticle massless birefringent Majorana fermions. Low energy effective models are constructed to understand the basic mathematical structures of equations of motion and physical properties for the gapless birefringent Majorana fermions. The equation of motion describing gapless birefringent Majorana fermions in condensed matter physics is the birefringence version of the original Majorana equation of real relativistic spin-1/2 fermions, with two different "speed of light". The birefringent Majorana fermions may be observed in superconductors or related interacting many-body systems as collective excitations due to the deep relationship between superconductors and Majorana fermions. In addition, birefringent Majorana fermions with mass in condensed matter physics are also valuable to investigate.

## ACKNOWLEDGMENTS

This work was supported by the Advanced Talents Incubation Program of the Hebei University (Grants No. 521000981423) and the high-performance computing center of Hebei University.

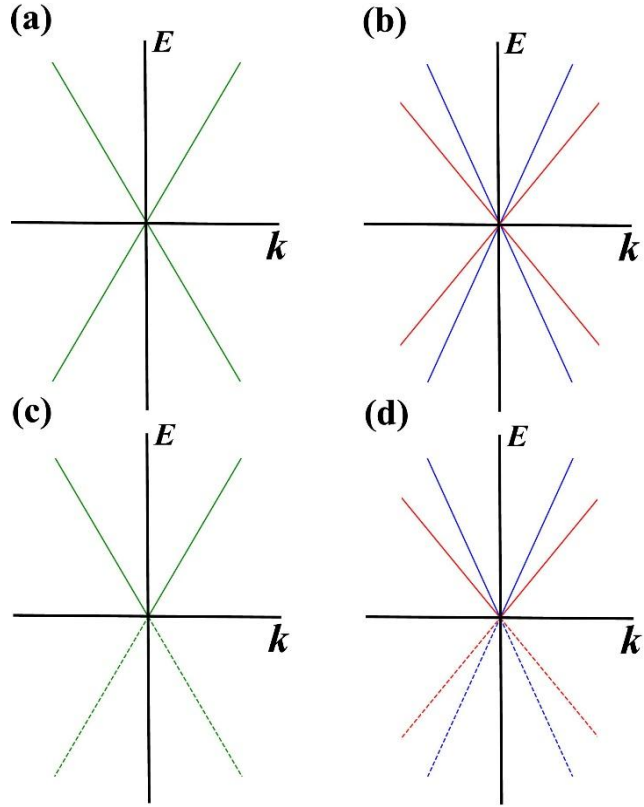

FIG. 1. Schematic comparison of (a) massless Dirac fermions, (b) massless birefringent Dirac fermions, (c) massless Majorana fermions and (d) massless birefringent Majorana fermions. The dashed lines show that the $E < 0$ part of the quasiparticle spectra are redundant for the Majorana fermions.

[1] A. H. Castro Neto, F. Guinea, N. M. R. Peres, K. S. Novoselov, and A. K. Geim, The electronic properties of graphene, Rev. Mod. Phys. **81**, 109 (2009).

[2] N. P. Armitage, E. J. Mele, and A. Vishwanath, Weyl and Dirac semimetals in three-dimensional solids, Rev. Mod. Phys. **90**, 015001 (2018).

[3] Z. Wang, Y. Sun, X.-Q. Chen, C. Franchini, G. Xu, H. Weng, X. Dai, and Z. Fang, Dirac semimetal and topological phase transitions inA3Bi (A=Na, K, Rb), Phys. Rev. B **85**, 195320 (2012).

[4] M. P. Kennett, N. Komeilizadeh, K. Kaveh, and P. M. Smith, Birefringent breakup of Dirac fermions on a square optical lattice, Phys. Rev. A **83**, 053636 (2011).

[5] C. Chen, S.-S. Wang, L. Liu, Z.-M. Yu, X.-L. Sheng, Z. Chen, and S. A. Yang, Ternary wurtzite CaAgBi materials family: A playground for essential and accidental, type-I and type-II Dirac fermions, Phys. Rev. Mat. **1**, 044201 (2017).

[6] H. Zhang, W. Huang, J.-W. Mei, and X.-Q. Shi, Influences of spin-orbit coupling on Fermi surfaces and Dirac cones in ferroelectriclike polar metals, Phys. Rev. B **99**, 195154 (2019).

[7] S. R. Elliott and M. Franz, Colloquium: Majorana fermions in nuclear, particle, and solid-state physics, Rev. Mod. Phys. **87**, 137 (2015).

[8] S. B. Chung and S. C. Zhang, Detecting the Majorana fermion surface state of 3He-B through spin relaxation, Phys. Rev. Lett. **103**, 235301 (2009).

[9] X.-L. Qi and S.-C. Zhang, Topological insulators and superconductors, Rev. Mod. Phys. **83**, 1057 (2011).

[10] C. Chamon, R. Jackiw, Y. Nishida, S. Y. Pi, and L. Santos, Quantizing Majorana fermions in a superconductor, Phys. Rev. B **81**, 224515 (2010).

[11] C.-X. Liu, X.-L. Qi, H. Zhang, X. Dai, Z. Fang, and S.-C. Zhang, Model Hamiltonian for topological insulators, Phys. Rev. B **82**, 045122 (2010).

[12] E. Majorana, 1937, Nuovo Cimento 5, 171 [Soryushiron Kenkyu 63,149 (1981)].